\documentclass[conference,10pt]{IEEEtran}

\IEEEoverridecommandlockouts

\usepackage[T1]{fontenc}
\usepackage{amssymb,amsmath}

\usepackage{balance}
\usepackage{algorithmic}
\usepackage[lined,ruled,linesnumbered]{algorithm2e}
\usepackage[]{graphicx}
\usepackage{xcolor}

\newcommand{\bW}{\boldsymbol{W}}
\newcommand{\bR}{\boldsymbol{R}}
\newcommand{\bx}{\boldsymbol{x}}

\newcommand{\bdelta}{\boldsymbol{\delta}}
\newcommand{\by}{\boldsymbol{y}}
\newcommand{\bt}{\boldsymbol{t}}
\newcommand{\bv}{\boldsymbol{v}}
\newcommand{\bd}{\boldsymbol{d}}
\newcommand{\bg}{\boldsymbol{g}}
\newcommand{\ba}{\boldsymbol{a}}
\newcommand{\bb}{\boldsymbol{b}}
\newcommand{\bu}{\boldsymbol{u}}
\newcommand{\bm}{\boldsymbol{m}}
\newcommand{\be}{\boldsymbol{e}}
\newcommand{\bz}{\boldsymbol{z}}
\newcommand{\bI}{\boldsymbol{I}}
\newcommand{\bL}{\boldsymbol{L}}
\newcommand{\bQ}{\boldsymbol{Q}}

\newcommand{\bone}{\boldsymbol{1}}

\def\R{\ensuremath{\mathrm{I\!R}}}

\usepackage{amsthm}

\newtheorem{proposition}{Proposition}

\begin{document}
\title{\huge Distributed Change Detection in Streaming Graph Signals\vspace{-2mm}}
\author{A. Ferrari, C. Richard and L. Verduci\\
Universit{\'e} C{\^o}te d'Azur\\ Observatoire de la C\^ote d'Azur, CNRS, France\thanks{This work was funded in part by the PIA program under its IDEX UCAJEDI project (ANR-15-IDEX-0001) and by ANR under grant ANR-19-CE48-0002.}}

\maketitle

\begin{abstract}
Detecting abrupt changes in streaming graph signals is relevant in a variety of applications ranging from energy and water supplies, to environmental monitoring. In this paper, we address this problem when anomalies activate localized groups of nodes in a network. We  introduce  an  online  change-point  detection algorithm,  which  is  fully  distributed  across  nodes  to  monitor large-scale dynamic networks. We analyze the detection statistics for controlling the probability of a global type 1 error. Finally we  illustrate  the detection and  localization performance with simulated data.
\end{abstract}

\section{Introduction}

Data generated by network-structured applications such as sensor networks, smart grids, and communication networks, usually lie on complex and irregular supports~\cite{chen2014multitask,chen2015diffusion,nassif2014multitask,nassif2016proximal}. These data need specific graph signal processing (GSP) tools to exploit their characteristics. As an extension of historical signal processing applied to data on graphs, GSP has received increasing attention in recent years~\cite{Shuman2013}. Ongoing researches include spectral analysis~\cite{marques2017stationary} and filtering~\cite{coutino2018advances}, to cite a few.

Detecting anomalous events on graph signals is relevant in a variety of applications ranging from surveillance of energy and water supplies, to environmental monitoring. The anomalies in these problems often tend to activate localized groups of nodes in the networks. The problem of deciding, based on noisy measurements at each node of graph, whether the underlying unknown graph signal is in a nominal state over the graph, or there exists a cluster of nodes with anomalous activation, was recently addressed with GSP tools~\cite{Sharpnack2016}. This work addresses  critical issues raised by theoretical contributions in the statistical literature~\cite{addario2010combinatorial,arias2011detection}. In particular, the generalized likelihood ratio (GLR) test is a natural solution for detecting anomalous clusters of activity in graphs. It however suffers from its computational complexity since it consists of scanning all well-connected clusters of nodes and testing them individually. The authors in~\cite{sharpnack2013changepoint} address this issue by incorporating, into the detection problem, the properties of the graph topology through its spectrum. They analyze the corresponding test statistics, which is based on a spectral measure of the combinatorial Laplacian, and show that it is indeed related to the problem of finding sparsest cuts. The same authors introduce another detector in~\cite{Sharpnack2016}, the Graph Fourier Scan Statistic (GFSS), which consists of a low-pass filter based on the graph Fourier transform. This detector appears as a relaxation of its initial form in~\cite{sharpnack2013changepoint} and, because of its particular expression, its statistical power can be precisely characterized. None of these works address the problem of detecting anomalous clusters of activity over time from streaming data at each node, nor do they implement the detectors in a fully distributed manner across nodes. 

In this paper, we address the problem of detecting abrupt changes activating localized groups of nodes, in a streaming graph signal. We introduce an online change-point detection algorithm, which is fully distributed across nodes to monitor large-scale dynamic networks. The remainder of this paper is organized as follows. Section~\ref{sec:problem-formulation} formulates the problem, and shows how it can be addressed with a centralized algorithm. Section~\ref{sec:distributed-algorithm} makes it scalable by devising an online distributed alternative. Section~\ref{sec:stat} is dedicated to its statistical analysis and section~\ref{sec:simul} to numerical experiments.  Finally, some concluding remarks are provided.

\section{Online change-point detection}
\label{sec:problem-formulation}

We consider an undirected graph $\mathcal{G}=\{V,E,\bW\}$ with $p$ vertices $V = \{1,\ldots,p\}$, $m$ edges $(i,j) \in E \subset V \times V$, and a $p\times p$ weighted adjacency matrix $\bW$ where $\bW_{i,j} = \bW_{j,i} \geq 0$ denotes the connection strength of $(i, j) \in  E$. 
 
The GFSS introduced in~\cite{Sharpnack2016} considers a single graph-signal measurement $\by\in\R^n$, i.e., a snapshot of the graph signal. It tests whether $\by$ is constant over all vertices, or there exists a cluster $C^\ast$ of \emph{well-connected} nodes where the average signal level differs from its complement $V\backslash C^\ast$. Compared to this paper, we consider here the sequential scenario where, at time instant $t$ we observe a graph signal $\by_t$. We denote by $\by_t(i)$ the noisy signal sample observed at time $t$ and vertex $i$. We shall assume that $\by_t = \bm_t + \be_t$, where $\bm_t(i)$ is the mean value at vertex $i$ and time instant $t$, and $\be_t$ is a noise  assumed temporally i.i.d. and spatially distributed as $\be_t\sim \mathcal{N}(\boldsymbol{0},\sigma^2\bI)$.

The objective is to detect an abrupt change of $\bm_t$ upon an unknown cluster $C^\ast$ at an unknown time $t_r$ (change-point):
\begin{equation}
    \label{eq:initial-problem}
	t<t_r:\; \bm_t = \bm,\qquad\;t\geq t_r:\ \bm_t = \bm + \bdelta
\end{equation}
where $\bdelta$ is non-zero only on $C^\ast$ with a constant value over this cluster. These assumptions differ from~\cite{Sharpnack2016} in that \eqref{eq:initial-problem} depends on time $t$ and $\bm$ is not necessarily constant over all vertices.

\subsection{GFSS algorithm}

Let $\bL$ be the normalized graph Laplacian of $\mathcal{G}$, and define by $\bu_i$, $i=1,\ldots,p$ the set of orthonormal eigenvectors of $\bL$ with $\mu_i$ the associated eigenvalues. 
Given a graph signal $\by$ measured on~$\mathcal{G}$, the GFSS is defined as:
\begin{align}
    & t_\text{GFSS}(\by) = \|\bg(\by)\|_2 
    \label{GFSS1}\\
    & \bg(\by) = \sum_{i=2}^p h^\ast(\mu_i)(\bu_i^\top \by)\bu_i 
    \label{GFSS2}
\end{align}
where $\bg(\by)$ is the graph-filtered signal in~\eqref{GFSS2}, and $h^\ast(\mu)$ is the frequency response of the filter defined as~\cite{Sharpnack2016}:
\begin{equation}
    h^\ast(\mu) = \min\left\{1, \sqrt{\frac{\gamma}{\mu}} \right\} ,\; \mu>0\label{filt_GFSS}
\end{equation}
where $\gamma >0$ is a tuning parameter.
See~\cite{Shuman2013,Tremblay2018,Ortega2018} for details on graph filtering.
In the case of a time-varying graph signal $\by_t$ as described in \eqref{eq:initial-problem}, $t_\text{GFSS}(\by_t)$ has proved to be inefficient in many situations. For instance, statistic $\bg(\by_t)$ does not vary much if $\bm_t$ entries increases and decreases simultaneously on two clusters at some instant $t_r$.

\subsection{Adaptive GFSS algorithm}

Comparing probability distributions that underlie data in a past and present interval has been proved to be an appealing tool for change-point detection (CPD). The most prominent methods are cumulative sum algorithms (CUSUM). They rely on parametric model assumptions that may not be met in practice. Non-parametric CPD algorithms were introduced to handle scenarios where no prior information on the data distribution and the nature of the change is available. Some~\cite{hawkins2010nonparametric} consist of an estimation phase of the nominal state followed by the detection phase. Other strategies perform online estimation of a time-varying model~\cite{yang2018sequential,xie2013change-point}. Finally, some methods are model-free in the sense that they compare a set of recent samples with samples that came before~\cite{kifer2004detecting,liu2013change-point,bouchikhi2018nonparametric,bouchikhi2019kernel}.

In this paper, we consider the adaptive strategy in~\cite{keriven2018newma} where two statistics are learned simultaneously using two distinct learning rates: $\bv_t$ with $\lambda$ and $\bv'_t$ with $\Lambda$, where $0<\lambda<\Lambda<1$:
\begin{equation}
	\bv_t = (1-\lambda)\bv_{t-1} + \lambda \by_t ,\quad
	\bv'_t = (1-\Lambda)\bv_{t-1} + \Lambda \by_t \label{bvy}
\end{equation}
When initialized with $\bv_{-1} = \bv'_{-1} = \boldsymbol{0}$, $\bv_t$ and $\bv'_t$ are both exponentially weighted averages of $\by_t$:
\begin{equation}
    \bv_t = \sum_{k=0}^t \lambda(1-\lambda)^{t-k}   \by_{k},\quad
    \bv'_t = \sum_{k=0}^t \Lambda (1-\Lambda)^{t-k} \by_{k}
\end{equation}
with forgetting factors $(1-\lambda)^{t-k}>(1-\Lambda)^{t-k}$. 

When $\by_t=\bm+\be_t$ and $t\rightarrow\infty$, $\bv_t\stackrel{a}{\sim}\mathcal{N}(\bm,\sigma^2\lambda\bI)$. Besides noise reduction by factor $\lambda$, the vector $\bv_t$ can be subtracted from the short-time average $\bv'_t$ in order to fit the assumptions in~\cite{Sharpnack2016}. First, when $t<t_r$, the mean of the signal of interest should be ideally constant over all vertices: here we have $\mathbb{E}[\bv'_t-\bv_t] \rightarrow \boldsymbol{0}$. Second, when $t\geq t_r$, a change appears on the mean of a cluster $C^\ast$ of well-connected vertices. This supports our interest in using $\bv_t-\bv'_t$ with GFSS for CPD.

\begin{figure}
	\centerline{\includegraphics[width=.6\columnwidth]{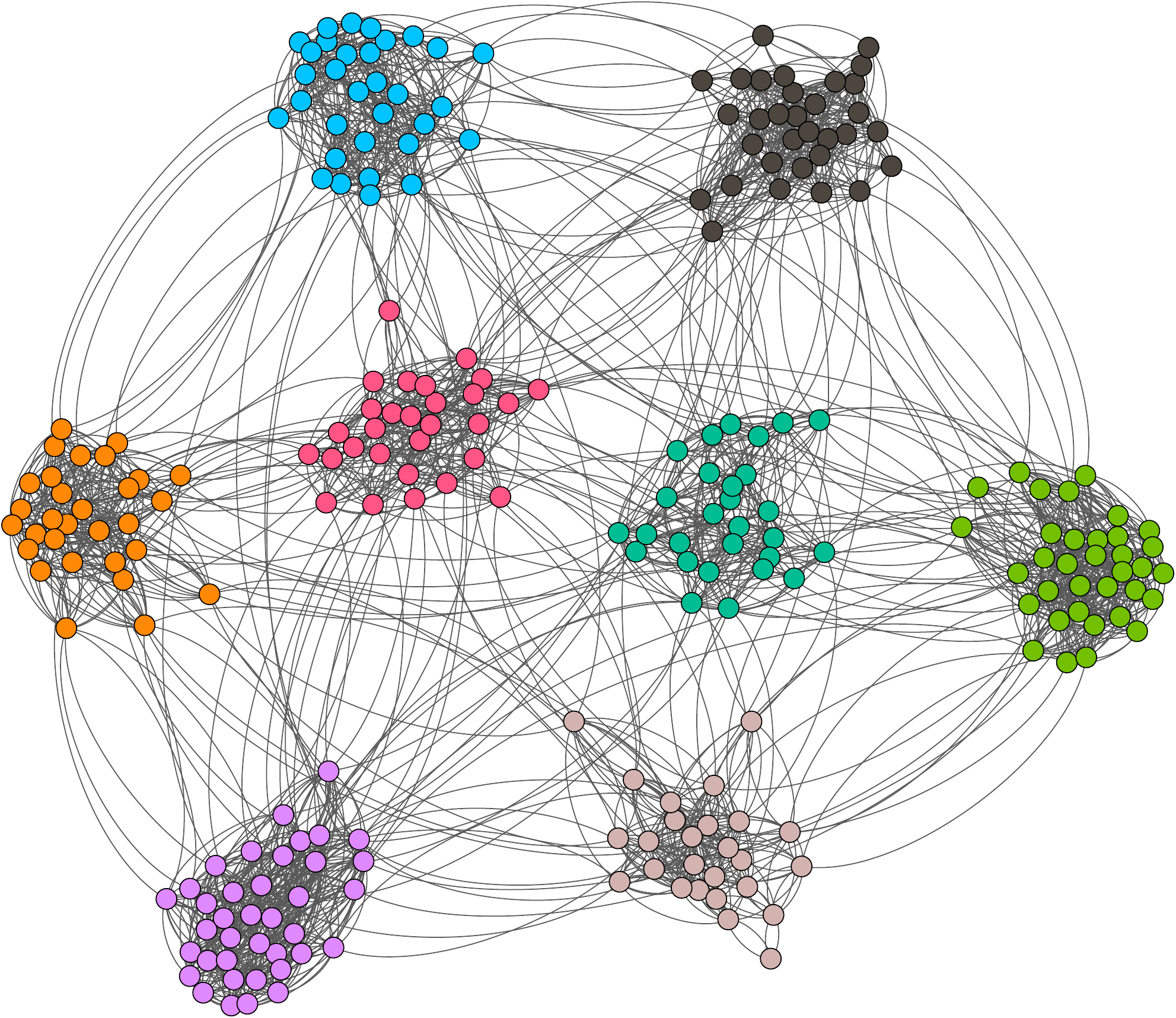}}
	\caption{Graph topology. The graph is unweighted and contains $p=250$ vertices and $m=2508$ edges. Clusters have been unfolded using \cite{Blondel2008} and colored for visualization.}
	\label{fig:aGFSS1}
\end{figure}

\begin{figure}
	\centerline{\includegraphics[width=.9\columnwidth]{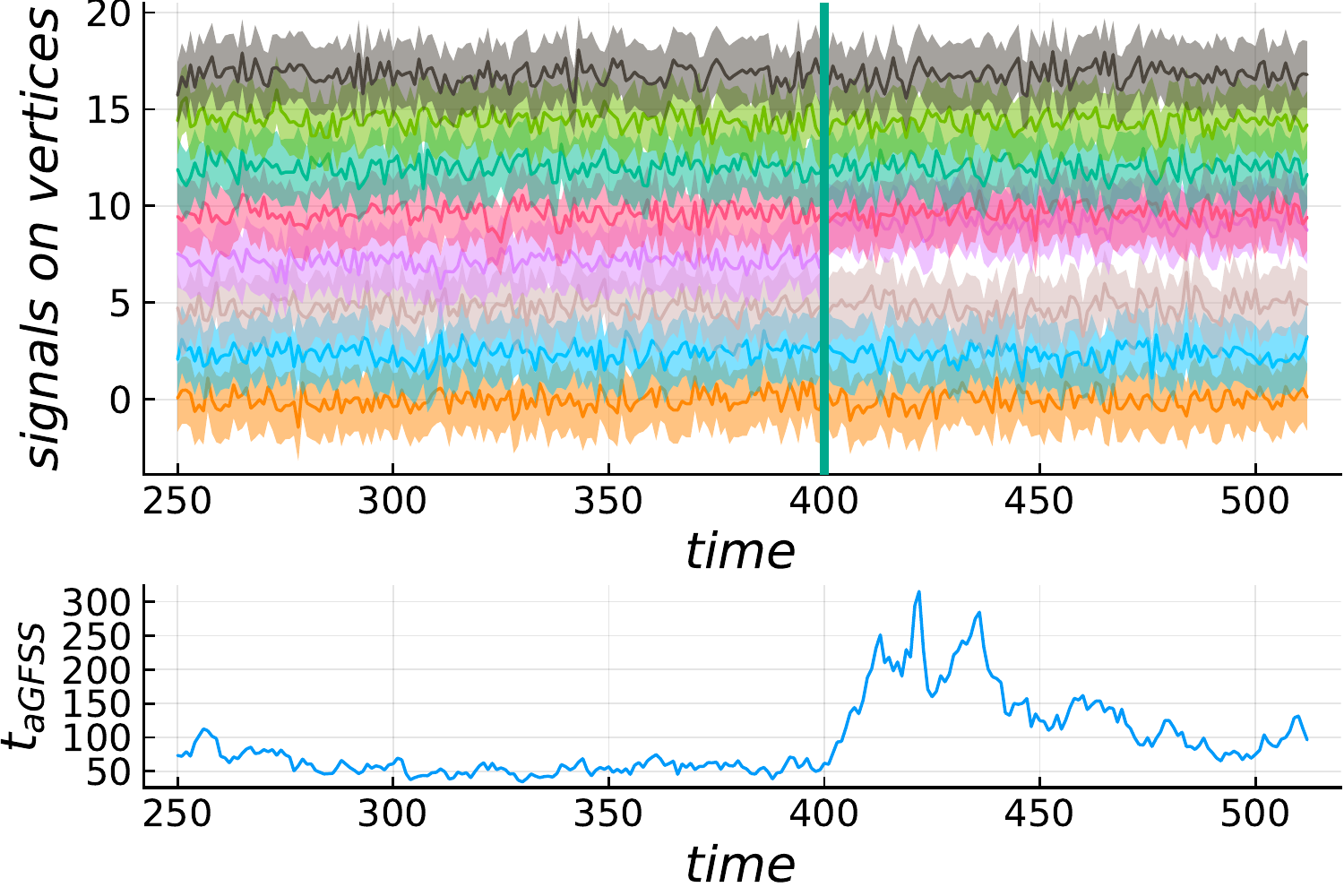}}
	\caption{Signals on vertices $\by_t(i)$ and test statistic $t_{aGFSS}$. The colors correspond to the colors of the clusters, see Fig. \ref{fig:aGFSS1}. The plots represent, for each cluster, the mean of the signal with a ribbon illustrating the noise standard deviation. An abrupt change occurs at $t=400$ on the ``magenta'' cluster. Parameters of the aGFSS were set to $\gamma=0.3$, $\lambda = 0.01$, $\Lambda=0.1$.}
	\label{fig:aGFSS2}
\end{figure}

The proposed Adaptive Graph Fourier Scan Statistics, denoted as aGFSS, is described in Alg.~\ref{aGFSS}. Since $\bg(\cdot)$ in~\eqref{GFSS2} is linear with respect to its argument, averaging can be equivalently performed on $\by_t$ or $\bz_t = \bg(\by_t)$ to calculate the output of the graph filter $\bg(\by_t)$. The performances of this algorithm are illustrated in Figs. \ref{fig:aGFSS1} and \ref{fig:aGFSS2}.

\begin{algorithm}
\begin{algorithmic}[1]
\STATE Learning rate $0<\lambda<\Lambda<1$, threshold $\xi$ 
\FOR{$t=1,2,\ldots$}
\STATE{$\bz_t = \bg(\by_t)$}
\STATE{$\bv_t = (1-\lambda)\bv_{t-1} + \lambda\bz_t$}
\STATE{$\bv'_t = (1-\Lambda)\bv_{t-1} + \Lambda\bz_t$}
\STATE{$t_\text{aGFSS} = \|\bv_t - \bv'_t\|_2$}
\IF {$t_\text{aGFSS}>\xi$}
	\STATE {flag $t$ as a change point on the graph}
\ENDIF
\ENDFOR
\end{algorithmic}
\caption{Adaptive GFSS}
\label{aGFSS}
\end{algorithm}

However, the aGFSS suffers from the following weaknesses:
\begin{enumerate}
\item Computation of $\bz_t$ in Step 3 is centralized. It requires at each time instant $t$ to transmit the signals at all vertices to a master node.
\item The test statistic $t_\text{aGFSS}$ in Step 6 is centralized and does not allow to localize the change spatially. When a change point is detected, a clustering algorithm must be run to find the subset of vertices where the change occurred.
\item The algorithm cannot adapt to a change in the graph topology. If a subset of edges is modified at a time instant, say $t_e$, the eigen-decomposition of the updated normalized graph Laplacian must be performed to be able to compute the graph filter output.
\end{enumerate}

\section{Distributed and adaptive GFSS}
\label{sec:distributed-algorithm}

A strategy to make Alg.~\ref{aGFSS} scalable consists of substituting the actual filter in Step 1 by a filter that can be distributed over the graph vertices as in~\cite{Shuman2011,Sandryhaila2014,Safavi2015,Segarra2015}. In contrast to these finite impulse response filters, an autoregressive moving average (ARMA) graph filter was proposed in \cite{Loukas2015, Isufi2017}. The parallel ARMA$_K$ graph filter is defined by:
\begin{align}
    &\bx_{\ell,t} = \psi_\ell \bL \bx_{\ell,t-1} + \varphi_\ell \by_t,\;\bx_{\ell,-1}=\boldsymbol{0},\; \forall \ell=1\ldots K, 
    \label{eq:armaK1}\\
    &\bz_t = \sum_{\ell=1}^K\bx_{\ell,t} + c \by_t
    \label{eq:armaK2}
\end{align}
The computation of the $i$-th entry of $\bL \bx_{\ell,t-1}$ by vertex $i$ only requires neighboring vertices to send it samples $\bx_{\ell, t-1}(j)$. The asymptotic properties of $\bz_t$ in Proposition 1 of Sec.~\ref{sec:stat} motivates the use of an ARMA$_K$ filter to approximate the graph filtering process used by the aGFSS. Practical problems related to the computation of parameters $c$ and  $(\psi_\ell, \varphi_\ell)$ for $\ell=1,\ldots, K$ are discussed in Sec.~\ref{sec:simul}. 

As noted above, the GFSS defined in \eqref{GFSS1}--\eqref{GFSS2} cannot localize the cluster where a change occurs. In order to get more insight into this statistics, it is important to recall the role played by the eigenvectors $\bu_i$ of the (normalized) graph Laplacian matrix $\bL$ in spectral clustering. Consider first the ideal case of a graph with $k>1$ connected components. The eigenvectors $\bu_i$, $i=1,\ldots,k$ of the Laplacian are the indicator vectors of the $k$ connected components of the graph, and $\bu_i^\top \by$ in~\eqref{GFSS2} is proportional to the sum of the signal samples of component $i$. Replacing $h^\ast(\mu_i)$ in~\eqref{GFSS2} by $h^\circ(\mu_i) = 1$ for $1\leq i\leq k$, and $0$ otherwise, $\bg(\by)$ assigns to each vertex $i$ of each component~$\ell$, the weighted sum of the signal samples $\by(j)$ at all vertices $j$ that belong to component $\ell$.  The number of components $k$ being unknown, $h^\ast(\mu_i)$ in~\eqref{filt_GFSS} then acts as a proxy of $h^\circ(\mu_i)$ that penalizes large numbers of components. If the components are connected by few edges, we can assume that this analysis is still approximately valid; see, e.g., \cite{vonLuxburg2007,Ng2002,Shi2000}.

This interpretation of the GFSS test statistics motivates the need to monitor, at each vertex $i$, the $i$-th entry of $\bg(\by)$ rather than monitoring $\|\bg(\by)\|_2$ globally. We shall denote by $\bg(\by)(i)$ the $i$-th entry of $\bg(\by)$. As an alternative to this fully local scheme, we shall consider the monitoring of each vertex~$i$ including its neighborhood $N_\mathcal{G}(i)$. Along the same line as the standard GFSS in~\eqref{GFSS1}, a possible strategy consists of testing the squared 2-norm $\sum_{k\in N_\mathcal{G}(i)}[\bg(\by)(k)]^2$. However, given~\eqref{eq:initial-problem} which assumes that changes take the form of a constant shift~$\bdelta$ in the mean at all vertices of a particular cluster, we shall also focus on the \emph{coherent sum}: $\sum_{k\in N_\mathcal{G}(i)}\bg(\by)(k)$. The rational behind these three test statistics is to localize the places where changes occur when they are detected. The use of an ARMA$_K$ filter allows to implement these strategies in a fully distributed manner as shown in~Alg.~\ref{daGFSS}. Note that a variation of the graph topology over time can be easily taken into account by substituting $\bL$ by $\bL_t$ in Step~3: as in ~\cite{Isufi2017}.

\begin{algorithm}
\begin{algorithmic}[1]
\STATE Learning rate $0<\lambda<\Lambda<1$, thresholds $\xi_i$, $i=1\ldots p$
\FOR{$t=1,2,\ldots$}
\FOR{$\ell=1,\ldots,K$}
\STATE{$\bx_{\ell,t} = \psi_\ell \bL \bx_{\ell,t-1} + \varphi_\ell \by_t$}
\ENDFOR
\STATE{$\bz_t = \sum_{\ell=1}^K\bx_{\ell,t} + c \by_t$}
\STATE{$\bv_t = (1-\lambda)\bv_{t-1} + \lambda\bz_t$}
\STATE{$\bv_t' = (1-\Lambda)\bv_{t-1}' + \Lambda\bz_t$}
\STATE{$\bd_t = \bv_{t}' -\bv_{t}$}
\STATE{$\forall i \in V$, $\bt_\text{daGFSS}(i) = \sum_{k\in N_\mathcal{G}(i) }\bd_t(k)$}
\IF {$\exists i \in V$: $|\bt_\text{daGFSS}(i)|>\xi_i$}
	\STATE {flag $t$ as a change point on the graph at vertex $i$}
\ENDIF
\ENDFOR
\end{algorithmic}
\caption{Distributed and Adaptive GFSS}
\label{daGFSS}
\end{algorithm}

\section{Statistical analysis and test settings} 
\label{sec:stat}

The aim of this section is to derive the statistical properties of $\bt_\text{daGFSS}(i)$ in order to set the threshold values $\xi_i$, $i\in V$. Using standard properties of linear dynamical systems, \cite{Anderson05}, results in \cite{Isufi2017} can be extended to the random case. Proofs are omitted due to the lack of space.

\begin{proposition}
	Consider the graph signal $\by_t=\bm+\be_t$ with $\be_t$ defined as in Sec.~I. If $\max_\ell\{|\psi_\ell|\} \rho(\bL)<1$ with $\rho(\cdot)$ the spectral radius of its matrix argument, then $\bz_t$ is asymptotically distributed as $\bz_t \stackrel{a}{\sim} \mathcal{N}(\bar{\bz}_\infty,\bQ_\infty)$ where  $\bar{\bz}_\infty$ is the graph signal $\bm$ graph-filtered by:
\begin{equation}
h(\mu) = c + \sum_{\ell = 1}^K\frac{\varphi_\ell}{1-\psi_\ell\mu}
\label{filtARMAK}
\end{equation}
and:
\begin{equation}
\bQ_\infty = \sum_{i=1}^p \kappa(\mu_i)\bu_i\bu_i^\top,\;
\kappa(\mu) = \sum_{\ell,\ell' =1}^K \frac{\sigma^2 \varphi_\ell\varphi_{\ell'}^\ast }{1 - \psi_\ell\psi_{\ell'}^\ast\mu^2}+\eta \label{cov_assympt}
\end{equation}
with $\eta=c\sigma^2(c +2\sum_{\ell = 1}^K\varphi_\ell)$.
\end{proposition}

\begin{proposition}
    Consider the graph signal $\by_t=\bm+\be_t$ with $\be_t$ defined as in Sec.~I. If $\max_\ell\{|\psi_\ell|\} \rho(\bL)<1$, then $\bd_t= \bv_{t}' -\bv_{t}$ is asymptotically distributed as $\bd_t \stackrel{a}{\sim} \mathcal{N}(\boldsymbol{0},\bR_\infty)$ with:
    \begin{equation}
	    \bR_\infty = \eta\bQ_\infty,\quad
	    \eta = \frac{\lambda}{2-\lambda} + \frac{\Lambda}{2-\Lambda} -\frac{2\lambda\Lambda}{\lambda+\Lambda-\lambda\Lambda}\label{covd}
    \end{equation}
\end{proposition}

These results are useful to set the Probability of False Alarm (PFA). If we focus on node $i$, $\bt_\text{daGFSS}(i)\stackrel{a}{\sim}\mathcal{N}(0,\sigma_{i}^2)$ where
\begin{equation}
\sigma_{i}^2 = \sum_{(k,\ell)\in N_\mathcal{G}(i) }\bR_\infty(k,\ell)
\end{equation}

Detecting abrupt changes over streaming graph signals by monitoring all $\bt_\text{daGFSS}(i)$, $i\in V$, is a multiple testing problem~\cite{Efron2012}. A classical approach for controlling a global Type I error $\alpha$ within this context is the False Discovery Rate (FDR)~\cite{Benjamini1995}, e.g., with the Benjamini-Hochberg procedure. This strategy however requires the ordering of all p-values of $\bt_\text{daGFSS}(i)$ for all $i$, which must be performed in a centralized manner. It is also important to note that the local test statistics are not independent. A solution may consist of decorrelating $\bt_\text{daGFSS}$, but this cannot be performed in a distributed manner.
For these reasons, we propose to consider a Bonferroni correction, i.e., setting the PFA at each node to $\alpha/p$. This leads to set in Alg.~\ref{daGFSS}:
\begin{equation}
	\xi_i = \sqrt{2}\sigma_i \text{erfc}^{-1}\left( \frac{\alpha}{p}\right)
\end{equation}

\section{Simulations} \label{sec:simul}

Approximating the GFSS filter~\eqref{filt_GFSS} by the ARMA$_K$ filter~\eqref{filtARMAK} is a necessary preliminary step in implementing Alg~\ref{daGFSS}. Starting the summation in \eqref{GFSS2} at $i=2$ is equivalent to set $h^\ast(0)=0$ in~\eqref{filt_GFSS}. We observed that this strong discontinuity deteriorates the approximation of $h^\ast(\mu)$ by the frequency response $h(\mu)$ in~\eqref{filtARMAK}. We propose to relax this constraint by setting $h^\ast(0)=1$. Note that if the standard Laplacian is used, we have $\bu_1=\bone$. Setting $h^\ast(0)=1$ then adds the same quantity $\sum_{k=1}^V\by(k)$ to each entry of $\bg(\by)$ and, consequently, does not affect the detection performance.

Given a filter order $K$, and similarly to the code provided in \cite{Loukascode}, the filter parameters can be computed by minimizing a quadratic loss. We considered minimizing:
\begin{equation}
\mathcal{J}(\ba, \bb)=  \sum_i \left[ B(x_i) - h(x_i)A(x_i)\right]^2 \label{}
\end{equation}
w.r.t. $(\ba,\bb)$ where $A(x) = 1+ \sum_{\ell=1}^K \ba(\ell) x^\ell=\prod_{\ell=1}^K(1-\psi_\ell x)$, and $B(x) = \sum_{\ell=0}^K \bb(\ell) x^\ell$, over a uniform grid $x_i$ on the interval $(0,2)$. According to Prop.~1, this quadratic problem must be solved subject to $|\psi_\ell| <\rho(\bL)^{-1}$ for all $\ell$. Note that these constraints are related to variables $\{\psi_\ell\}$ while the optimization problem is solved w.r.t. $(\ba,\bb)$. To address this issue, these constraints are relaxed to the following constraints which are linear w.r.t. $\ba$: $|A(x_i)|< \beta$ for all $x_i$ on the grid, with $\beta$ a parameter to be set by the user. Finally, initial variables $c$ and $\{(\phi_\ell,\psi_\ell)\}$ were estimated from $(\ba,\bb)$ by a partial fraction expansion of  $B(x)/A(x)$.

\begin{figure}
	\centerline{\includegraphics[width=.6\columnwidth]{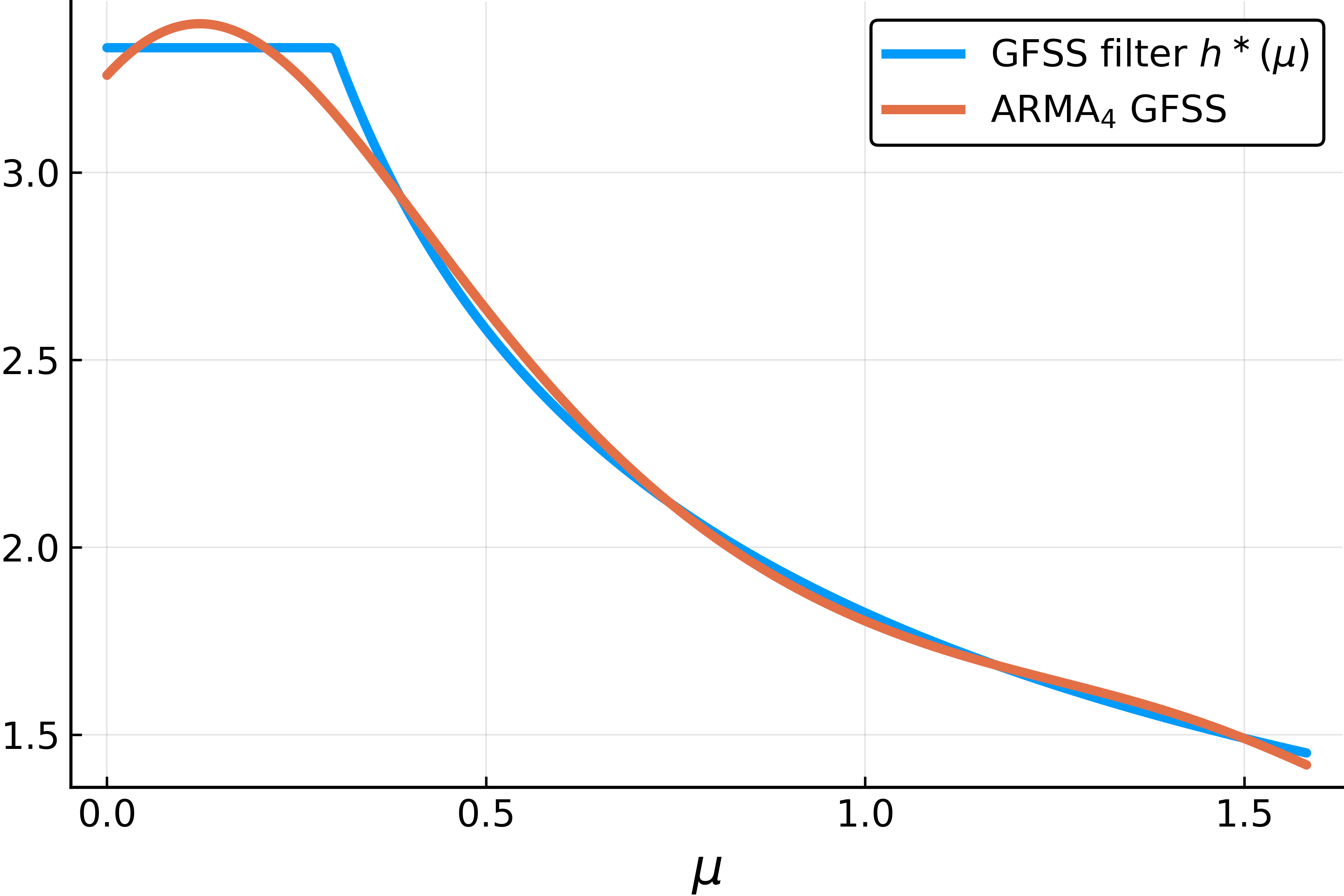}}
	\caption{ARMA$_4$ approximation of the GFSS filter}
	\label{fig:filts}
\end{figure}

\begin{figure}
	\centerline{\includegraphics[width=.5\columnwidth]{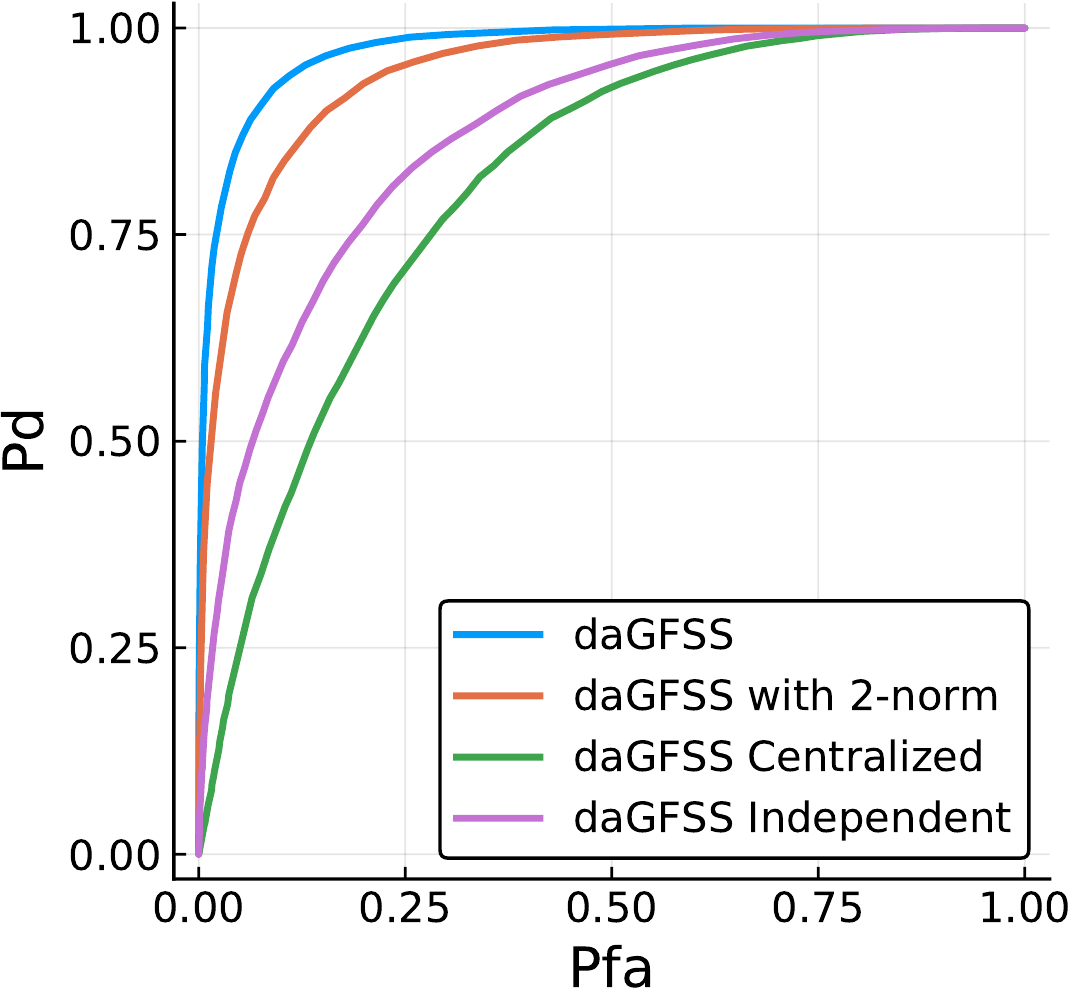}}
	\caption{ROC curve. ``dAGFSS'': Alg. \ref{daGFSS},  ``dAGFSS with 2-norm'': Alg. \ref{daGFSS} with test statistics based on the 2-norm, ``dAGFSS centralized'': test statistic $\|\bd_t\|_2$, ``daGFSS independent'': test at each vertex with $|\bd_t(i)|$.}
	\label{fig:roc}
\end{figure}

\begin{figure}
	\vspace*{-1cm}\centerline{\includegraphics[width=.5\columnwidth]{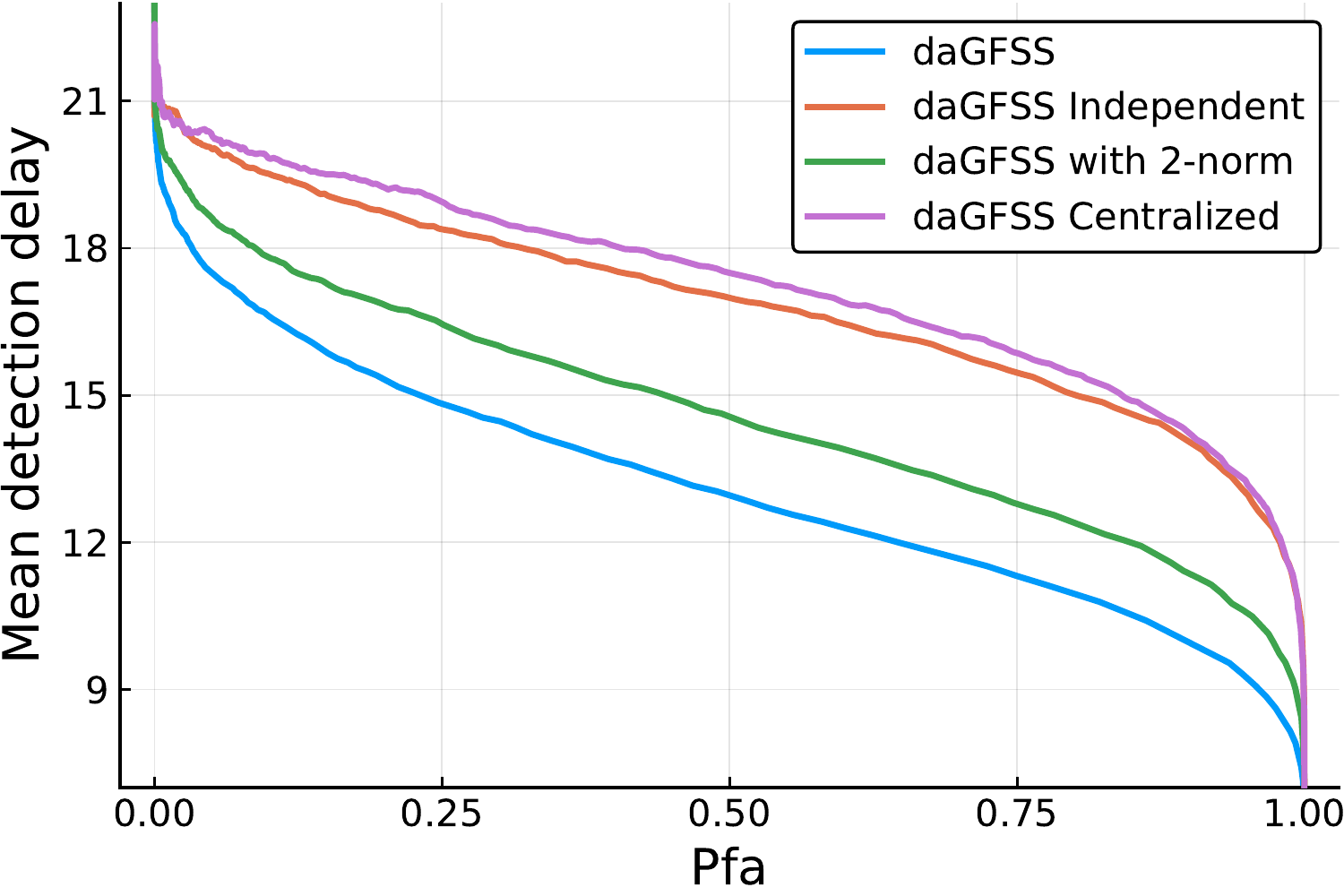}}
	\caption{Detection delay. See Fig. \ref{fig:roc} for the labels.}
	\label{fig:delay}
\end{figure}

\begin{figure}
	\centerline{\includegraphics[width=.5\columnwidth]{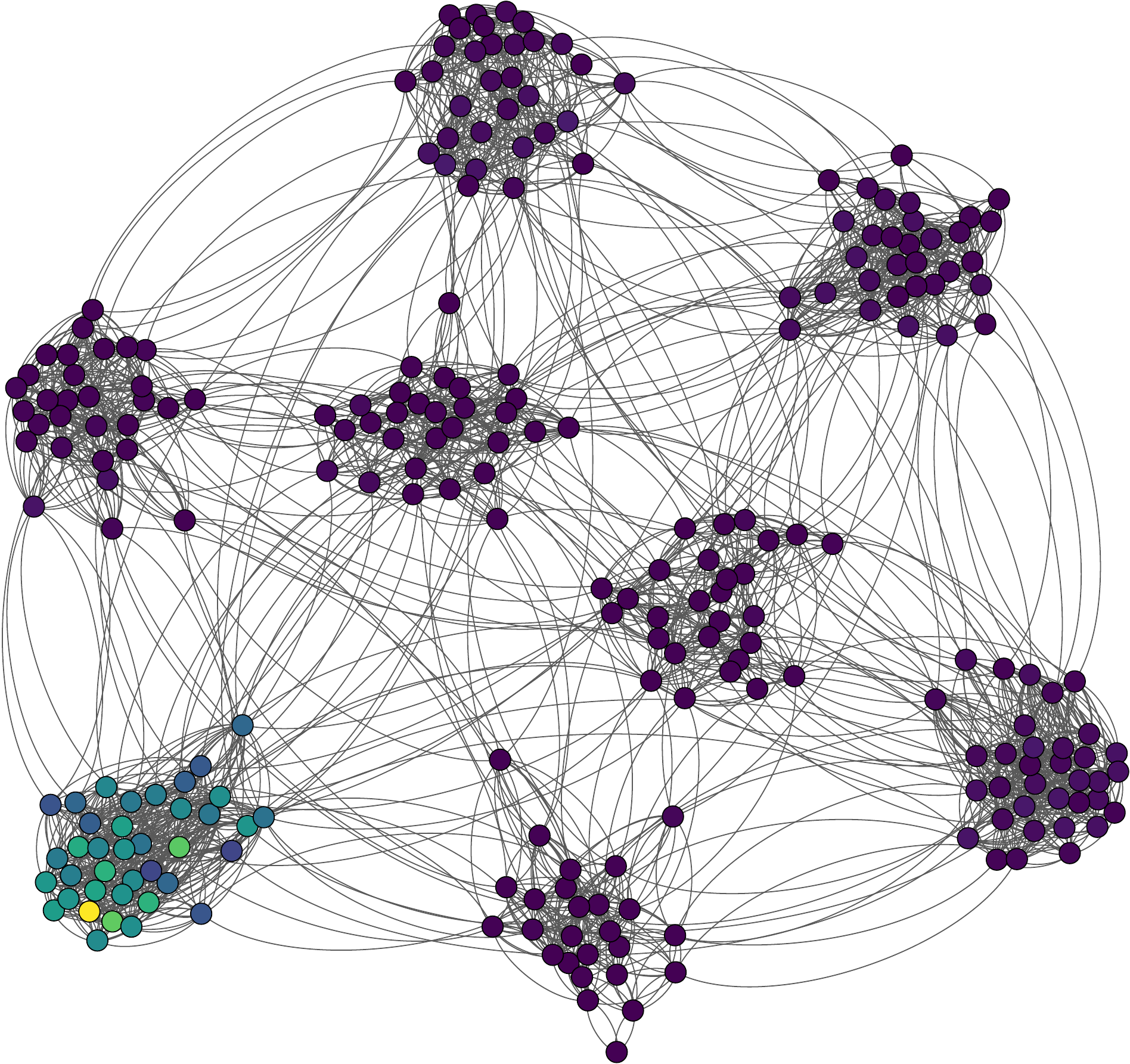}\quad \includegraphics[width=.07\columnwidth]{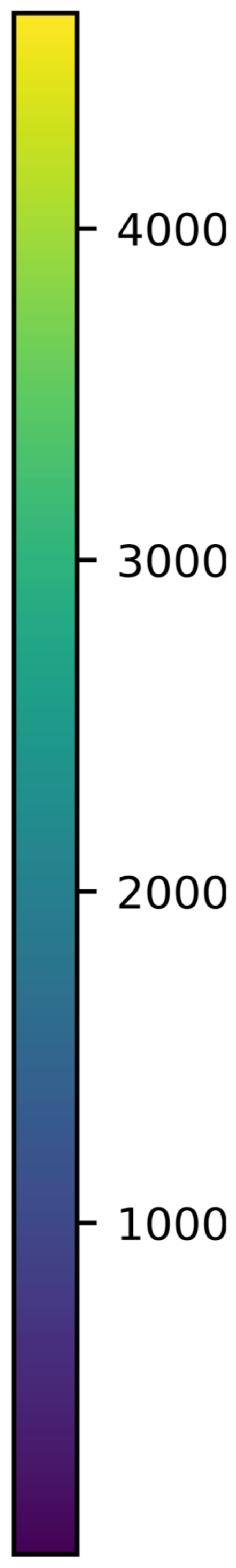}}
	\caption{Test statistic $\bt_\text{daGFSS}(i)$ for $t=t_r+20$.}
	\label{fig:testongraph}
\end{figure}

Figure~\ref{fig:filts} illustrates the approximation $h(\mu)$ of $h^\ast(\mu)$ with $\gamma=0.3$, for $K=4$ and $\beta=0.1$. We subsequently checked that the resulting filter $h(\mu)$ is stable.
The graph used for the simulations is represented in Fig. \ref{fig:aGFSS1}. The signal $\bm$ is defined as $\bm(i)=k$ for all vertices $i$ in cluster $k=0\ldots 7$. The noise variance is $\sigma^2=7$. The total number of samples is 512 and the abrupt change at $t_r=400$ consisted of $\bdelta(i)=0.5$ for the vertices $i$ in the pink-colored cluster, see Fig. \ref{fig:aGFSS1}. Figure \ref{fig:roc} represents the ROC curves of the different detectors considered in this paper, and Fig. \ref{fig:delay} provides the corresponding detection delays. Finally, Fig.~\ref{fig:testongraph} provides the test statistic $\bt_\text{daGFSS}(i)$ at each node after the change point, illustrating the ability of the algorithm to localize the cluster where the change occurred. Julia code is available at \texttt{github.com/andferrari/daGFSS}.

\section{Conclusion and perspectives}

In this paper, we introduced an online change-point detection algorithm, which is fully distributed across nodes to monitor large-scale dynamic networks. We illustrated its detection and localization performance with simulated data. Perspectives of this work include change-point detection over graphs with time-varying topology.

\clearpage
\balance
\bibliographystyle{IEEEtran}
\bibliography{MyBiblio}
\end{document}